\documentclass[preprintnumbers,amsmath,amssymbm,preprint]{revtex4}
\usepackage{epsfig}
\usepackage{graphicx}
\usepackage{amssymb}

\begin{document}
\title{Introducing the inverse hoop conjecture for black holes}
\author{Shahar Hod}
\affiliation{The Ruppin Academic Center, Emeq Hefer 40250, Israel}
\affiliation{ } \affiliation{The Hadassah Institute, Jerusalem
91010, Israel}
\date{\today}

\begin{abstract}
\ \ \ It is conjectured that stationary black holes are characterized by the inverse hoop relation 
${\cal A}\leq {\cal C}^2/\pi$, where ${\cal A}$ and ${\cal C}$ are respectively the black-hole surface area and 
the circumference length 
of the smallest ring that can engulf the black-hole horizon in every direction. We explicitly prove that 
generic Kerr-Newman-(anti)-de Sitter black holes conform to this conjectured area-circumference relation. 
\end{abstract}
\bigskip
\maketitle

\section{Introduction}

The isoperimetric inequality \cite{Iso} in a two-dimensional Euclidean space states that 
the area ${\cal A}$ of a connected domain 
is bounded from above by the simple relation 
\begin{equation}\label{Eq1}
{\cal A}\leq {\cal C}^2/4\pi\  ,
\end{equation}
 where ${\cal C}$ is the circumference length of the two-dimensional domain. 
 The equality in (\ref{Eq1}) may be attained by an engulfing circular ring. 

On the other hand, the area ${\cal A}$ of a deformed (or wrinkled) two-dimensional patch which is 
embedded in a three-dimensional space can violate the area-circumference relation (\ref{Eq1}) \cite{Notecir12}. 
Likewise, the surface area of a $(3+1)$-dimensional black hole 
may in principle grow unboundedly with respect to its (squared) circumference length. 

Intriguingly, however, it is well known that black holes in three spatial dimensions behave in many 
respects as two-dimensional objects. In particular, a black hole is 
characterized by a thermodynamic entropy \cite{Bekent,Hawent} which 
is proportional to its two-dimensional surface area (and not to its effective 
volume). 
One can therefore expect that, in analogy with the two-dimensional relation (\ref{Eq1}), 
the surface area of a black hole may 
be bounded from above by a quadratic function of its circumference length. 

The main goal of the present compact paper is to raise the {\it inverse hoop conjecture}, according to which 
the surface areas of all stationary $(3+1)$-dimensional 
black holes are bounded from above by the simple functional 
relation 
\begin{equation}\label{Eq2}
{\cal A}\leq {\cal C}^2_{\text{s}}/\pi\  ,
\end{equation}
where ${\cal C}_{\text{s}}$ is the circumference length 
of the smallest ring that can engulf the black-hole horizon in all azimuthal directions \cite{NoteThor,Thorne,Noterec,Hodst,Peng}.

\section{The inverse hoop conjecture in charged and spinning Kerr-Newman-(anti)-de Sitter black-hole spacetimes}

A Kerr-Newman-(anti)-de Sitter black-hole spacetime of 
mass $M$, angular momentum $J\equiv Ma$, electric charge $Q$, and cosmological
constant $\Lambda$ is characterized by the curved line element \cite{CarLam,Suz,Notebl,Noteunit}
\begin{equation}\label{Eq3}
ds^2=-{{\Delta_r}\over{\rho^2}}\Big({{dt}\over{I}}-a\sin^2\theta{{d\phi}\over{I}}\Big)^2
+{{\Delta_{\theta}\sin^2\theta}\over{\rho^2}}\Big[{{adt}\over{I}}-(r^2+a^2){{d\phi}\over{I}}\Big]^2
+\rho^2\Big({{dr^2}\over{\Delta_r}}+{{d\theta^2}\over{\Delta_{\theta}}}\Big)\  ,
\end{equation}
where the metric functions $\Delta_r,\Delta_{\theta},\rho$, and $I$ are given by the 
functional expressions \cite{CarLam,Suz}
\begin{equation}\label{Eq4}
\Delta_r\equiv r^2-2Mr+Q^2+a^2-{1\over3}\Lambda r^2(r^2+a^2)\  ,
\end{equation}
\begin{equation}\label{Eq5}
\Delta_{\theta}\equiv 1+{1\over3}\Lambda a^2\cos^2\theta\  ,
\end{equation}
\begin{equation}\label{Eq6}
\rho^2\equiv r^2+a^2\cos^2\theta\  ,
\end{equation}
and
\begin{equation}\label{Eq7}
I\equiv 1+{1\over3}\Lambda a^2\  .
\end{equation}

Asymptotically flat Kerr-Newman black holes are characterized by the simple relation $\Lambda=0$, 
whereas non-asymptotically flat 
Kerr-Newman-de Sitter and Kerr-Newman-anti-de Sitter black-hole spacetimes are 
characterized respectively by the relations $\Lambda>0$ and $\Lambda<0$. 
The horizon radii of the black-hole spacetime (\ref{Eq3}) are 
determined by the roots of the radial metric function $\Delta_r(r)$ \cite{CarLam,Suz,Notenoh}. In particular, 
\begin{equation}\label{Eq8}
\Delta_r(r_+)=0\  ,
\end{equation}
where $r_+$ is the radius of the black-hole event horizon. 

From Eqs. (\ref{Eq3}) and (\ref{Eq8}) one finds the compact expressions 
\begin{equation}\label{Eq9}
{\cal C}_{\text{eq}}=2\pi{{r^2_++a^2}\over{r_+ I}}\
\end{equation}
and 
\begin{equation}\label{Eq10}
{\cal A}=4\pi{{r^2_++a^2}\over{I}}\
\end{equation}
for the equatorial circumference 
and the horizon surface area of the 
Kerr-Newman-(anti)-de Sitter black hole.

Interestingly, from Eqs. (\ref{Eq9}) and (\ref{Eq10}) one finds the compact dimensionless ratio
\begin{equation}\label{Eq11}
{\cal H}(M,Q,a,\Lambda)\equiv {{\pi{\cal A}}\over{{\cal C}^2_{\text{eq}}}}={{Ir^2_+}\over{r^2_++a^2}}\
\end{equation}
for generic Kerr-Newman-(anti)-de Sitter black holes \cite{Noteaps}. The conjectured 
inverse hoop relation asserts that stationary $(3+1)$-dimensional black holes are characterized by 
the simple relation 
\begin{equation}\label{Eq12}
{\cal H}\leq1\  .
\end{equation}

Taking cognizance of Eqs. (\ref{Eq7}) and (\ref{Eq11}), one finds 
that asymptotically flat Kerr-Newman black holes (with $\Lambda=0$ and therefore $I=1$) 
and Kerr-Newman-anti-de Sitter black holes (with $\Lambda<0$ and therefore $I<1$) conform to the 
inverse hoop relation (\ref{Eq12}). It is easy to show that Kerr-Newman-de Sitter black holes (with $\Lambda>0$) 
are characterized by the relation $\Lambda r^2_+\leq1$ \cite{Notelam} and therefore also 
respect the inverse hoop relation (\ref{Eq12}).

\section{Summary and discussion}

The famous Thorne hoop conjecture \cite{Thorne} asserts 
that black-hole spacetimes of suitably defined mass ${\cal M}$ are characterized by the 
relation ${\cal M}\geq{\cal C}/4\pi$. Since there are many different definitions of mass (energy) in curved 
spacetimes, it is natural to ask: what is the exact physical meaning 
of the mass (energy) term ${\cal M}$ in the hoop relation? 
To the best of our knowledge, in his original work Thorne \cite{Thorne} has not provided a specific 
definition for the mass term ${\cal M}$ in the intriguing hoop conjecture. 
 
In the present compact paper we have explicitly demonstrated that if the mass term ${\cal M}$ is 
interpreted as the irreducible mass ${\cal M}_{\text{irr}}$ of the black hole, then 
generic Kerr-Newman-(anti)-de Sitter black-hole spacetimes conform 
to the {\it inverse} hoop relation 
\begin{equation}\label{Eq13}
{\cal M}_{\text{irr}}\leq{\cal C}_{\text{s}}/4\pi\  .
\end{equation}

Taking cognizance of the fact that the irreducible mass of a black hole is related to its horizon  
surface area ${\cal A}$ by the simple relation 
\begin{equation}\label{Eq14}
{\cal M}_{\text{irr}}\equiv\sqrt{{\cal A}/16\pi}\  ,
\end{equation}
one realizes that the inverse hoop relation (\ref{Eq13}) 
is a statement about the geometric properties of the black-hole horizon, 
bounding its surface area in terms of the squared circumference of the smallest ring that can 
engulf the horizon in every direction:
\begin{equation}\label{Eq15}
{\cal A}\leq {\cal C}^2_{\text{s}}/\pi\  .
\end{equation}
If true, the conjectured inverse hoop relation (\ref{Eq15}) implies that the black-hole surface area 
cannot be unboundedly wrinkled \cite{Notenbh}. 

Finally, it is worth noting that there is an important  numerical evidence \cite{East} for the 
validity of the inverse hoop conjecture (\ref{Eq15}) in non-stationary (dynamical) black-hole spacetimes. 
In particular, in a very interesting work, East \cite{East} has studied numerically the full non-linear 
gravitational collapse of self-gravitating spheroidal matter configurations. Remarkably, it has been explicitly 
demonstrated in \cite{East} that, in accord with the weak cosmic censorship conjecture \cite{Penw}, 
the final state of the collapse is a black hole. Interestingly, the initially distorted dynamically formed horizons 
obtained in \cite{East} are characterized by damped oscillations between being prolate and oblate 
(see Figure 1. of \cite{East}).

Intriguingly, and most importantly for our analysis, the numerical data presented in \cite{East} 
(see, in particular, Figure 1. of \cite{East}) reveals the fact that, 
within the bounds of the numerical accuracy \cite{Notepri}, 
the dynamically formed black holes presented in \cite{East} are 
characterized by the relation
\begin{equation}\label{Eq16}
{{\text{max}\{{\cal C}_{\text{eq}}(t),{\cal C}_{\text{p}}(t)\}}\over{4\pi{\cal M}_{\text{irr}}}}\geq1\  ,
\end{equation}
where ${\cal C}_{\text{eq}}$ and ${\cal C}_{\text{p}}$ are respectively the time-dependent (oscillating) equatorial and 
polar circumferences of the non-stationary black-hole horizons. Thus, the dynamically formed black holes 
presented in \cite{East} seem to respect 
the conjectured inverse hoop relation (\ref{Eq15}). 

\bigskip
\noindent {\bf ACKNOWLEDGMENTS}

This research is supported by the Carmel Science Foundation. I would like to thank Professor W. E. East for 
sharing with me his interesting numerical data. I would also
like to thank Yael Oren, Arbel M. Ongo, Ayelet B. Lata, and Alona B.
Tea for stimulating discussions.


\newpage

\begin{thebibliography}{99}

\bibitem{Iso} See R. Osserman, Bulletin of the American Mathematical Society, {\bf 84}, 6 (1978) and references therein.

\bibitem{Notecir12} For a two-dimensional patch embedded in a three-dimensional space, the parameter ${\cal C}$ 
may be defined as the circumference length of the smallest ring that can engulf that patch in all directions.

\bibitem{Bekent} J. D. Bekenstein, Phys. Rev. D {\bf 7}, 2333 (1973).

\bibitem{Hawent} S. W. Hawking, Commun. Math. Phys. {\bf 43}, 199 (1975).

\bibitem{Thorne} K. S. Thorne, in {\it Magic without Magic: John Archibald Wheeler},
edited by J. Klauder (Freeman, San Francisco, 1972).

\bibitem{NoteThor} The mathematically compact and physically influential 
Thorne hoop conjecture \cite{Thorne} asserts 
that black-hole spacetimes are characterized by the 
relation ${\cal M}\geq{\cal C}/4\pi$. It is worth noting that the exact physical meaning 
of the mass (energy) term ${\cal M}$ in the hoop relation 
has not been specified in the pioneering work of Thorne \cite{Thorne}. 
In the present compact paper we shall explicitly demonstrate that if the mass term ${\cal M}$ is 
interpreted as the irreducible mass ${\cal M}_{\text{irr}}\equiv\sqrt{{\cal A}/16\pi}$ of the black hole, then 
generic Kerr-Newman-(anti)-de Sitter black-hole spacetimes conform 
to the {\it inverse} hoop relation ${\cal M}_{\text{irr}}\leq{\cal C}/4\pi$.

\bibitem{Noterec} See \cite{Hodst,Peng} and references therein for recent studies of the 
Thorne hoop conjecture \cite{Thorne}.

\bibitem{Hodst} S. Hod, The Euro. Phys. Jour. C {\bf 78}, 1013 (2018) [arXiv:1903.09786].

\bibitem{Peng} Y. Peng, The Euro. Phys. Jour. C {\bf 79}, 943 (2019).

\bibitem{CarLam} B. Carter in {\it Les Astres Occlus}, edited by B. DeWitt, C. M. DeWitt,
(Gordon and Breach, New York, 1973).

\bibitem{Suz} H. Suzuki, E. Takasugi and H. Umetsu, Prog. Theor. Phys. {\bf 100}, 491
(1998).

\bibitem{Noteunit} We use natural units in which $G=c=\hbar=1$.

\bibitem{Notebl} Here we use the familiar Boyer-Lindquist spacetime
coordinates $(t,r,\theta,\phi)$ \cite{CarLam,Suz}.

\bibitem{Notenoh} Note that generic Kerr-Newman and Kerr-Newman-anti-de Sitter black-hole spacetimes 
are characterized by two (Cauchy and event) horizons, whereas generic 
Kerr-Newman-de Sitter black-hole spacetimes are characterized by three (Cauchy, event, and cosmological) 
horizons. 


\bibitem{Noteaps} Note that spinless ($a=0$) black holes saturate the inverse hoop conjecture (\ref{Eq2}) 
[see Eqs. (\ref{Eq7}) and (\ref{Eq11})].

\bibitem{Notelam} Note that an extremal Schwarzschild-de Sitter black hole whose event horizon 
coincides with the cosmological horizon is characterized by the simple dimensionless relation $\Lambda r^2_+=1$.

\bibitem{Notenbh} It is worth emphasizing that the conjectured inverse hoop relation (\ref{Eq15}) 
is expected to be valid only for black holes. In particular, it is straightforward to imagine non-black hole objects 
that violate the area-circumference relation (\ref{Eq15}). For example, a moon-like object whose surface is covered 
with craters can violate the relation (\ref{Eq15}). Likewise, a non-black hole Coronavirus-like object, 
whose surface is covered with spikes, can violate the area-circumference relation (\ref{Eq15}).  

\bibitem{East} W. E. East, Phys. Rev. Lett. {\bf 122}, 231103 (2019).

\bibitem{Penw} R. Penrose, Riv. Nuovo Cim. {\bf 1}, 252 (1969).

\bibitem{Notepri} W. E. East, Private communication. 

\end{thebibliography}
\end{document}